\documentclass[11pt]{article}

\usepackage{amssymb,amsfonts,amsmath,amsthm}
\usepackage{longtable}
\usepackage{graphicx,textgreek}
\usepackage{amsthm}
\usepackage{booktabs}
\usepackage{authblk}

\usepackage{environ}
\NewEnviron{myequation}{%
\begin{equation}
\scalebox{0.9}{$\BODY$}
\end{equation}
}

\begin{document}

\title{Bifurcations and multi-stability in a model of cytokine-mediated autoimmunity}
\author{F. Fatehi$^{\rm 1}$, Y.N. Kyrychko$^{\rm 1}$, R. Molchanov$^{\rm 2}$, K.B. Blyuss$^{\rm 1}$\footnote{Corresponding author: K.Blyuss@sussex.ac.uk}}
\affil{$^{\rm 1}$ Department of Mathematics, University of Sussex, Falmer, Brighton BN1 9QH, UK}

\affil{$^{\rm 2}$ Dnipropetrovsk Medical Academy of the Ministry of Health of Ukraine, Dnipro 49044, Ukraine}

\maketitle

\begin{abstract}
This paper investigates the dynamics of immune response and autoimmunity with particular emphasis on the role of regulatory T cells (Tregs), T cells with different activation thresholds, and cytokines in mediating T cell activity. Analysis of the steady states yields parameter regions corresponding to regimes of normal clearance of viral infection, chronic infection, or autoimmune behaviour, and the boundaries of stability and bifurcations of relevant steady states are found in terms of system parameters. Numerical simulations are performed to illustrate different dynamical scenarios, and to identify basins of attraction of different steady states and periodic solutions, highlighting the important role played by the initial conditions in determining the outcome of immune interactions.
\end{abstract}

\section{Introduction}
\label{intro}

Autoimmune disease is a pathological condition characterised by the failure of the immune system to efficiently discriminate between self-antigens and foreign antigens, resulting in unwanted destruction of healthy organ cells. In the case of normal functioning, recognition of foreign epitopes 
presented on antigen presenting cells (APCs) to T lymphocytes results in proliferation and effector function from T cells, while cross-reactivity between epitopes leads to the possibility of T cell response against self-antigens \cite{Mason98,And00}. T cells with high level of self-reactivity are removed from the system by two different mechanisms: central tolerance and peripheral tolerance. Central tolerance is associated with the removal of autoreactive T cells during their development in the thymus, while the peripheral tolerance is usually controlled by regulatory T cells \cite{Wing06}. One should note that it is important for autoreactive T cells to be present in the periphery to maintain effective T cell repertoire through generation of new peripheral T cells, and the stimulus produced by the healthy cells would normally not be sufficient to trigger activation of autoreactive T cells.

Clinical observations suggest that autoimmune disease is usually focused in a specific organ or part of the body, such as central nervous system in multiple sclerosis, retina in the case of uveitis, or pancreatic $\beta$-cells in insulin-dependent diabetes mellitus type-1 \cite{Kerr08,Prat02,San10}. 
Significant efforts have been made to pinpoint causes of autoimmune disease, and a large number of contributing factors have been identified, which include genetic predisposition, age, previous immune challenges, exposure to pathogens etc. \cite{RB15}. One should note that even in the presence of genetic predisposition \cite{Caf08,Li08,Guil11}, further environmental triggers are required to initiate the onset of autoimmune disease, with infections being one of the main contributors \cite{Ger12,Mal13}. Spontaneous autoimmunity has been associated with the dysregulation of immune response against Epstein-Barr virus in patients with multiple sclerosis \cite{Corr06,munz,DD01}, while infections with Coxsackie viruses are associated with type-1 diabetes \cite{erco,Hor98}. A number of distinct mechanisms have been identified that explain how an infection of the host with a pathogen can subsequently trigger the onset of autoimmune disease. These mechanisms include bystander activation \cite{fuji3} and molecular mimicry \cite{von,erco}, which is particularly important in the context of autoimmunity caused by viral infections.

In terms of mathematical modelling of immune response and possible onset of autoimmunity, some of the early models analysed interactions between regulatory and effector T cells without investigating specific causes of autoimmunity, but instead focusing on T cell vaccination \cite{sege}. Borghans and De Boer \cite{borg1} and Borghans {et al.}~\cite{borg2} showed how autoimmune dynamics, that they defined as above-threshold oscillations in the number of autoreactive cells, can appear in such models. L\'eon {\it et al.}~\cite{leon1,leon2,leon3} have studied interactions between different T cells, and how they can affect regulation of immune response and control of autoimmunity. Carneiro {\it et al.}~\cite{carn} have presented on overview of that work and compared two possible mechanisms of immune self-tolerance that are either based on control by specific regulatory T cells, or result from tuning of T cell activation thresholds. Iawmi {\it et al.}~\cite{iwam1,iwam2} have analysed a model of immune response to a viral infection with an emphasis on explicitly including the virus population, and the effects of different forms of the growth function for susceptible cells on autoimmune dynamics. Despite this model's ability to demonstrate the emergence of autoimmunity, since it does not allow for a viral expansion, it cannot support a regime of normal viral clearance. Alexander and Wahl~\cite{alex} have focused on how interactions of professional APCs with effector and regulatory T cells can control autoimmune response. Burroughs {\it et al.}~\cite{burr1,Burr11b} have demonstrated how autoimmunity can arise through bystander activation mediated by cytokines. An excellent overview of some of the latest development in mathematical modelling of autoimmune disease can be found in a special issue on ``Theories and modelling of autoimmunity" \cite{Bern15}.

Since T cells are know to be fundamental for the dynamics of autoimmunity, several different methodologies have been proposed for the analysis of various roles they play in coordinating immune response. Experimental evidence suggests that a major component in controlling autoimmune behaviour is provided by regulatory T cells, which are activated by autoantigens and act to shut down immune responses \cite{Saka04,Jos12,cort}, while
impairment in the function of regulatory T cells results in autoimmune disease \cite{Font03,Khat03}. To model this process, Alexander and Wahl \cite{alex} and Burroughs {\it et al.}~\cite{burr1,Burr11b} have explicitly included a separate compartment for regulatory T cells that are activated by autoantigens and suppress the activity of autoreactive T cells. Another theoretical approach supported by experimental evidence is the idea that T cells have the capacity to adjust their activation threshold for response to stimulation by autoantigens depending on various environmental conditions or endogenous stochastic variation, which allows them to perform a variety of different immune functions. The associated framework of {\it tunable activation thresholds} was proposed for analysis of thymic \cite{Gsinger96} and peripheral T cell dynamics \cite{gros,GPaul00}, and has been subsequently used to analyse differences in activation/response thresholds that are dependent on the activation state of the T cell \cite{Bonn05}. van den Berg and Rand~\cite{berg} and  Scherer {\it et al.}~\cite{Scher04} have developed and analysed stochastic models for tuning of activation thresholds. The importance of tuning lies in the fact that it provides an effective mechanism for improving sensitivity and specificity of T cell signalling in a noisy environment \cite{Fein08,George05}, and both murine and human experiments have confirmed that activation of T cells can indeed dynamically change during their circulation \cite{Bit02,Nich00,Roe11,Stef02}. It is noteworthy that the need for activation thresholds for T cells can be derived directly from the first principles of signal detection theory \cite{Noest00}.

To model the dynamics of immune response to a viral infection and possible onset of autoimmunity, Blyuss and Nicholson \cite{blyu12,blyu15} have proposed and analysed a mathematical model that includes two types of T cells with different activation thresholds and allows for a biologically realistic situation where infection and autoimmune response occur in different organs of the host. Depending on parameter values, this model can exhibit the regime of normal viral clearance, a chronic infection, and an autoimmune state represented by endogenous oscillations in cell populations, associated with episodes of high viral production followed by long periods of quiescence. Such behaviour, associated in the clinical practice with relapses and remissions, has been observed in a number of autoimmune diseases, such as MS, autoimmune thyroid disease, and uveitis \cite{Bezra95,Davies97,Nyla12}. Despite its successes, this model has several limitations. One of those is the fact that the periodic oscillations in the model are only possible when the amount of free virus and the number of infected cells are also exhibiting oscillations, while in laboratory and clinical situations, one rather observes a situation where autoimmunity follows full clearance of the initial infection. Another issue is that this model does not exhibit bi-stability, which could explain clinical observations suggesting that patients with very similar parameters of immune response can have significantly different course and outcome of the infection.

Bi-stability between different dynamical states is an extremely important property of various biological systems, and it has already been studied in a variety of contexts, including neural networks \cite{Bao12,lai2016,Oli97}, gene regulatory networks \cite{Huang05,lai2018,MacAr09}, within-cell dynamics of RNA interference \cite{Neof17}, and immune dynamics \cite{Bly09,iwam1}. In immunology, bifurcations and multi-stability have been studied in the context of T cell differentiation \cite{Yates00,Leb16}, including the role of cytokines \cite{Lee14}. Ngina {\it et al.}~\cite{Ngina17} have investigated the dynamics of HIV from the perspective of interactions between HIV virions and CD8$^+$ T cells, and  identified a region of bi-stability associated with the backward bifurcation, where the system can reach either a virion-free or endemic equilibrium depending on the initial conditions. This observation has profound implications for developing an effective anti-retroviral therapy. Due an important role played by immune response in mediating the onset and development of cancer, a number of researchers have investigated bifurcations and possible multi-stability that can arise in these interactions. Piotrowska~\cite{Piotr16} has analysed a model of immune response to malignant tumours with an emphasis on the role of time delay associated with developing immune response. This model was shown to exhibit bi-stability, with the dynamics being determined by the initial size of the tumour.  Li and Levine~\cite{Li17} have investigated a cytokine-mediated bi-stability between immune-promoting and immune-suppressing states in the model of cancer-related inflammation. Anderson {\it et al.}~\cite{And15} have looked into interactions between cytokines and CD4$^+$ T cells for the purpose of immunotherapy. The bi-stability was shown to correspond to two simultaneously present levels at which the tumor can stabilise depending on the initial conditions.

In the specific context of autoimmunity, L\'eon {\it et al.}~\cite{leon1} have highlighted the importance of bi-stability between steady states with high populations of either regulatory, or effector T cells for effective representation of the adoptive transfer of tolerance. Roy {\it et al.}~\cite{roy2014} have developed a general kinetic model to capture the role of vitamin D in immunomodulatory responses, and they demonstrated that vitamin D extends the region of bi-stability, thus allowing immune regulation to be more robust with respect to changes in pathogenic stimulation. Baker {\it et al.}~\cite{Bak13} have analysed the dynamics of immune response during rheumatoid arthritis with particular emphasis on the effects of cytokines on bi-stability and treatment. Rapin {\it et al.}~\cite{rapin2011} have proposed a simple model of autoimmunity that displays a bi-stability between stable steady states corresponding to a healthy state and autoimmunity. The authors have shown how the system can be switched back to the healthy steady state by immunotherapy aimed at destabilising an autoimmune steady state.

In this paper, we will show how inclusion of regulatory T cells and the cytokine mediating T cell activity can allow one to overcome above-mentioned difficulties and provide a more realistic representation of various regimes in the dynamics of immune response. In the next section we will introduce the model and discuss its basic properties. Section \ref{sec:3} contains systematic analysis of all steady states, including conditions for their feasibility and stability. In Section \ref{sec:4} we perform extensive bifurcation analysis of the model and illustrate various types of behaviour that can be exhibited by the system depending on parameters and initial conditions, which includes identification of basins of attraction of various states. The paper concludes in Section \ref{sec:5} with the discussion of results.

\section{Model derivation}
\label{sec:2}

To analyse the dynamics of immune response to infection and possible onset of autoimmunity, we use an approach similar to some of the earlier models of immune response \cite{blyu12,blyu15,woda,nowa}. The underlying idea is the mechanism of molecular mimicry, where immune response against an infection can lead to a breakdown of immune tolerance due to cross-reaction with one or more self-antigens that share some of their immunological characteristics with a pathogen \cite{erco,von}. Experimental evidence suggests that while antibodies are important in a wider picture of immune response to viral infections, within the context of autoimmunity, B cells can be dispensable, so that autoimmune disease can develop even in their absence \cite{wolf96}. Moreover, it has been shown in some studies that the development of antibodies can itself depend on prior interactions of T cells with a pathogen \cite{wu10}. Hence, in this paper we rather focus on the role of T cells and associated cytokines.

We consider a situation where both infection and autoimmune response are targeting the same organ of the body, and the population of healthy cell in this organ is denoted by $A(t)$. These cells are assumed to follow logistic growth with the proliferation rate $r$ and the carrying capacity $N$ in the absence of infection or autoimmune response, as is common in models of viral dynamics \cite{iwam1,pere}. At the same time, one should be mindful of the fact that different functional forms of the growth of healthy cells can also have an effect on autoimmune dynamics, as has been shown by Iwami {\it et al.}~\cite{iwam1,iwam2}. 

During a viral infection, some number of healthy cells become infected by free virus particles, at which point they move to the compartment of infected cells, denoted by $F(t)$. After a certain period of time, these infected cells will be producing virions, or free virus particles, $V(t)$ at a rate $k$, and the rate of natural clearance of virions is denoted by $c$. These virions then go on to infect other as yet uninfected cells at a rate $\beta$, which is an effective rate incorporating time constants associated with various biological processes, such as the movement of virions, cell entry, and an eclipse phase, during which the cells are infected but are not yet recognised as such by the immune system.

In terms of immune dynamics, T cell response originates in the lymph nodes. Stimulation of na\"ive T cells results in their proliferation, differentiation into activated T cells, and subsequent migration to the infected tissue. Once activated, T cells bearing the CD8$^+$ receptor become cytotoxic T cells that are able to destroy infected cells, whereas if they have a CD4$^+$ receptor, they turn into helper T cells \cite{abbas,Kim2007}. Tregs perform an important role of suppressing the autoreactive T cells,
\begin{figure}[h]
	\centering
	\includegraphics[width=0.93\linewidth]{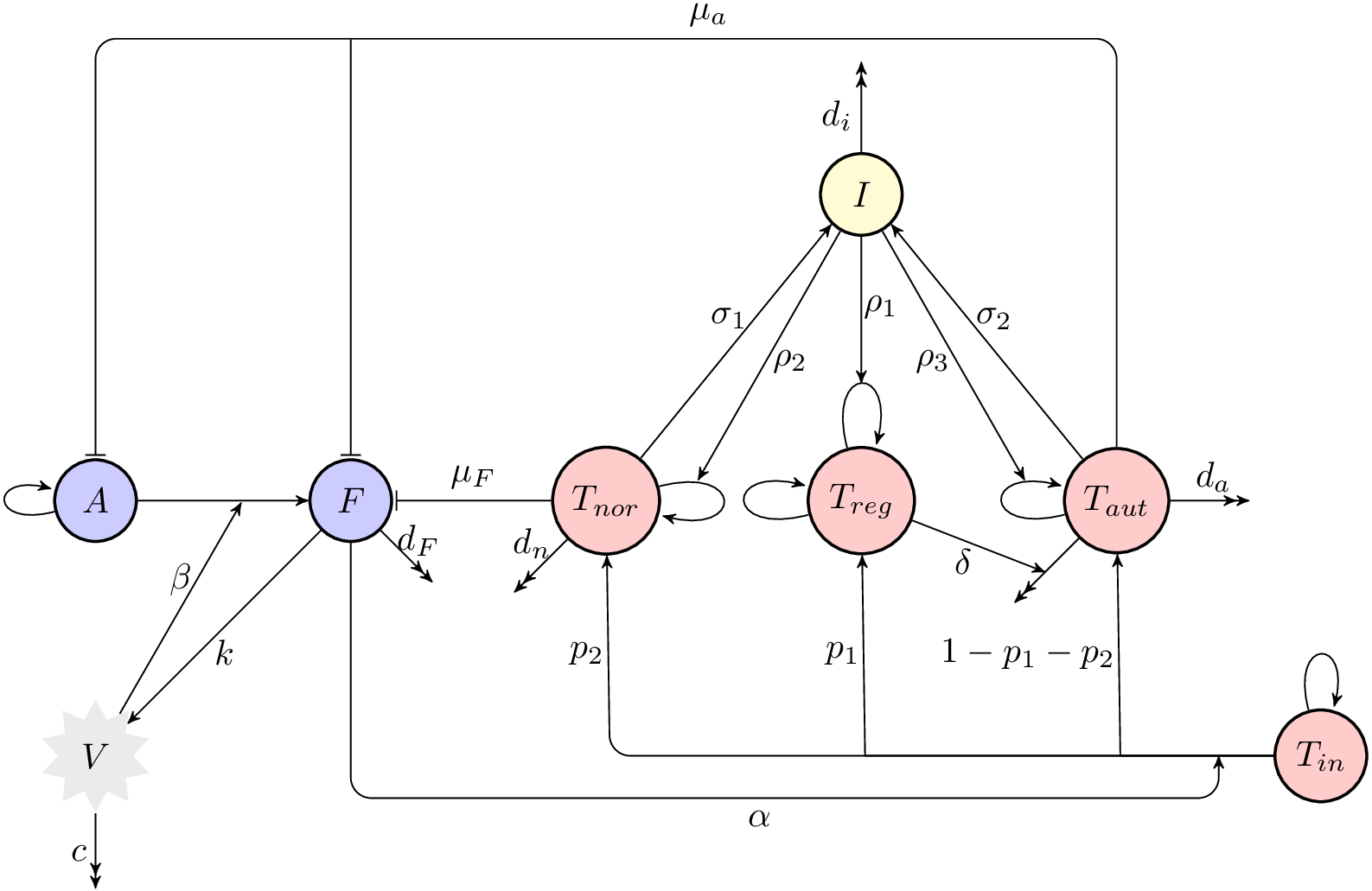}
	\caption{A schematic diagram of immune response to an infection. Blue circles indicate host cells (uninfected and infected cells), red circles denote different T cells (na\"ive, regulatory, normal activated, and autoreactive T cells), yellow circle shows cytokines (interleukin 2), and grey indicates virus particles (virions). Single arrow-headed and bar-headed lines indicate, respectively, production/proliferation and destroying of one cell population from/by another. Double arrows indicate natural clearance.}\label{dia}
\end{figure}
\noindent and are a part of CD4$^+$ T cell population \cite{Saka1995,janeway2005,Walker2005}. Since in this paper we are trying to understand self and non-self discrimination mechanisms of the immune response, we consider two populations of na\"ive CD8$^+$ T cells that respond to self-antigens and foreign antigens, while focusing on one population of CD4$^+$ T cells representing regulatory T cells. Kim {\it et al.}~\cite{Kim2007} have considered a situation where each population of na\"ive T cells is maintained at a certain level supported by homeostasis in the absence of infection. Burroughs {\it et al.}~\cite{burr2,burr1} and Segel {\it et al.}~\cite{sege} in their models have instead considered a constant influx of new T cells from the thymus. In this model, for simplicity, we consider a single population of na\"ive T cells which includes Tregs, foreign-reactive and self-reactive T cells, and, similarly to earlier work, the population of these na\"ive T cells is assumed to be maintained at a certain level by homeostasis \cite{blyu12,iwam1,iwam2,pere}. It is thus assumed that in the absence of infection, these cells are produced at a constant rate $\lambda_{in}$, and they die at a rate $d_{in}$. Once activated, these cells differentiate into either regulatory T cells, whose main role is the control of immune response against self- and foreign antigens \cite{abbas,cort}, as well as prevention of autoimmune disease \cite{cort,alex,carn,dank}, or effector cells that are able to eliminate infected cells. We denote by $\alpha$ the rate at which na\"ive T cells are activated. Since $T_{in}$ includes different kinds of na\"ive T cells, it is assumed that a constant proportion $p_1$ of them will develop into regulatory T cells $T_{reg}$, and a proportion $p_2$ will become normal activated T cells $T_{nor}$ that are able to recognise infected cells expressing foreign antigen and destroy these cells at rate $\mu_F$. The remaining proportion $(1-p_1-p_2)$ of T cells will become autoreactive cells $T_{aut}$ with a lower threshold for activation by healthy cells, hence, they will be destroying both infected and healthy cells at rate $\mu_a$. Unlike the work by Blyuss and Nicholson \cite{blyu12,blyu15}, Burroughs {\it et al.}~\cite{burr1} and Kim {\it et al.}~\cite{Kim2007} have not explicitly modelled the production of autoreactive T cells from normal activated T cells, and in the present model we also do not include this feature, as the model already accounts for the influx of each population of T cells directly to the tissue.

Similarly to other models of autoimmune dynamics \cite{burr1,burr2,balt}, regulatory T cells in our model are assumed to have their own homeostatic mechanism, and they are assumed to be produced at constant rate $\lambda_r$ and die at rate $d_r$. One of the main effects of regulatory T cells is to suppress the proliferation of autoreactive T cells. Part of this suppression occurs through the inhibition of interleukin 2 (IL-2) by T cells \cite{Thornton1998,Shevach2001}. Moreover, there is evidence for both cell-to-cell inhibition, and soluble mediators such as IL-10 and TNF-\textbeta \hspace{0.1cm}\cite{burr2,burr1,Saka04,Kim2007}. There is some experimental evidence suggesting that the suppression by Tregs is antigen-specific \cite{Yu2005,Tanchot2004,Tang2006}, which implies that Tregs are able to discriminate between T cells responding to self-antigens and T cells responding to foreign antigen \cite{cort}. L\'eon {\it et al.}~\cite{leon1} and Carneiro {\it et al.}~\cite{Carneiro2007} have proposed a model that considers antigen-specific suppression by Tregs, thus endowing Tregs with a mechanism for self/non-self discrimination. Baecher-Allan {\it et al.}~\cite{Baecher-Allan2002} have proposed a model for the T cell receptor (TCR) signal strength, where Tregs suppress the activation of autoreactive T cells, while the T cells reactive to foreign antigen are refractory to the suppression. Thus, in this paper we only consider direct suppression of autoreactive T cells by Tregs, which is assumed to occur at rate $\delta$, and, unlike the work by Burroughs {\it et al.}~\cite{burr1} and Kim {\it et al.}~\cite{Kim2007}, we are assuming Tregs do not suppress the normal activated T cells $T_{nor}$. Among various cytokines involved in the process of immune response, a particularly important role is played by IL-2, to be denoted by $I(t)$, which is an essential factor in the growth of T cells. Whilst this cytokine promotes the growth of both regulatory and effector T cells, regulatory T cells do not secrete IL-2 \cite{abbas,burr1,burr2}. Therefore, in this model we assume that $T_{nor}$ and $T_{aut}$ produce IL-2 at rates $\sigma_1$ and $\sigma_2$. On the other hand, whilst regulatory T cells do not produce IL-2, similar to other T cells they need this cytokine for their activation and proliferation \cite{Fat18a,Fat18b}. Thus, we assume that IL-2 promotes proliferation of $T_{reg}$, $T_{nor}$ and $T_{aut}$ at rates $\rho_1$, $\rho_2$ and $\rho_3$, respectively. Although it is also possible to include in the model inhibition of IL-2 by T cells \cite{burr2,Fat18b}, our analysis shows that this would not qualitatively change the behaviour of the model. 

With the above assumptions, the model for dynamics of immune response to a viral infection with account for Tregs, T cells with different activation thresholds and IL-2, as illustrated in Fig.~\ref{dia}, takes the form
\begin{equation}
\begin{array}{l}
\dfrac{dA}{dt}=rA\left(1-\dfrac{A}{N}\right)-\beta AV-\mu_a T_{aut}A,\\\\
\dfrac{dF}{dt}=\beta AV-d_FF-\mu_FT_{nor}F-\mu_a T_{aut}F,\\\\
\dfrac{dT_{in}}{dt}=\lambda_{in}-d_{in}T_{in}-\alpha T_{in}F,\\\\
\dfrac{dT_{reg}}{dt}=\lambda_r-d_rT_{reg}+p_1\alpha T_{in}F+\rho_1 IT_{reg},\\\\
\dfrac{dT_{nor}}{dt}=p_2\alpha T_{in}F-d_nT_{nor}+\rho_2 IT_{nor},\\\\
\dfrac{dT_{aut}}{dt}=(1-p_1-p_2)\alpha T_{in}F-d_aT_{aut}-\delta T_{reg}T_{aut}+\rho_3 IT_{aut},\\\\
\dfrac{dI}{dt}=\sigma_1T_{nor}+\sigma_2T_{aut}-d_iI,\\\\
\dfrac{dV}{dt}=kF-cV,
\end{array}
\end{equation}
with $0\leq p_1+p_2\leq 1$. Introducing non-dimensional variables
\begin{align*}
&\hat{t}=rt,\quad A=N\hat{A},\quad F=N\hat{F},\quad T_{in}=\dfrac{\lambda_{in}}{d_{in}}\hat{T}_{in},\quad T_{reg}=\dfrac{\lambda_{in}}{d_{in}}\hat{T}_{reg},\\ &T_{nor}=\dfrac{\lambda_{in}}{d_{in}}\hat{T}_{nor},\quad T_{aut}=\dfrac{\lambda_{in}}{d_{in}}\hat{T}_{aut},\quad I=\dfrac{\lambda_{in}}{d_{in}}\hat{I},\quad V=N\hat{V},
\end{align*}
yields a rescaled model
\begin{equation}\label{resc_model}
\begin{array}{l}
\dfrac{dA}{dt}=A\left(1-A\right)-\beta AV- \mu_aT_{aut}A,\\\\
\dfrac{dF}{dt}=\beta AV- d_FF-\mu_FT_{nor}F-\mu_aT_{aut}F,\\\\
\dfrac{dT_{in}}{dt}=d_{in}\left(1-T_{in}\right)-\alpha T_{in}F,\\\\
\dfrac{dT_{reg}}{dt}=\lambda_r-d_rT_{reg}+p_1\alpha T_{in} F+\rho_1IT_{reg},\\\\
\dfrac{dT_{nor}}{dt}=p_2\alpha T_{in}F-d_nT_{nor}+ \rho_2 IT_{nor},\\\\
\dfrac{dT_{aut}}{dt}=(1-p_1-p_2)\alpha T_{in}F-d_aT_{aut}- \delta T_{reg}T_{aut}+\rho_3IT_{aut},\\\\
\dfrac{dI}{dt}=\sigma_1T_{nor}+\sigma_2T_{aut}-d_iI,\\\\
\dfrac{dV}{dt}=kF-cV,
\end{array}
\end{equation}
where
\begin{align*}
&\hat{\beta}=\dfrac{\beta N}{r},\quad \hat{\mu}_a=\dfrac{\mu_a\lambda_{in}}{rd_{in}},\quad \hat{d}_F=\dfrac{d_F}{r},\quad \hat{\mu}_F=\dfrac{\mu_F\lambda_{in}}{rd_{in}},\quad \hat{d}_{in}=\dfrac{d_{in}}{r},\\
&\hat{\alpha}=\dfrac{\alpha N}{r}, \quad \hat{\lambda}_r=\dfrac{\lambda_rd_{in}}{\lambda_{in}r},\quad \hat{d}_n=\dfrac{d_n}{r}, \quad \hat{d}_a=\dfrac{d_a}{r}, \quad \hat{\rho}_i=\dfrac{\rho_i \lambda_{in}}{rd_{in}}, \quad i=1,2,3,\\
&\hat{\delta}=\dfrac{\delta \lambda_{in}}{r d_{in}},\quad \hat{\sigma}_1=\dfrac{\sigma_1}{r},\quad \hat{\sigma}_2=\dfrac{\sigma_2}{r},\quad \hat{d}_i=\dfrac{d_i}{r},\quad \hat{k}=\dfrac{k}{r},\quad \hat{c}=\dfrac{c}{r},\quad \hat{d}_r=\dfrac{d_r}{r},
\end{align*}
and all hats in variables and parameters have been dropped for simplicity of notation. The model~(\ref{resc_model}) is clearly well-posed, i.e. its solutions remain non-negative for $t\geq 0$ for any non-negative initial conditions.

\section{Steady states and their stability}
\label{sec:3}
As a first step in the analysis of model~(\ref{resc_model}), we look at its steady states
\[
S^{\ast}=\left(A^{\ast}, F^{\ast}, T^{\ast}_{in}, T^{\ast}_{reg}, T^{\ast}_{nor}, T^{\ast}_{aut}, I^{\ast},V^{\ast}\right),
\]
that can be found by equating to zero the right-hand sides of equations~(\ref{resc_model}) and solving the resulting system of algebraic equations. High dimensionality of the system~(\ref{resc_model}) results in a large number of possible steady states, so we now systematically study all of them. First, we consider a situation where at a steady state, there is no free virus population, i.e. $V^{\ast}=0$, which immediately implies $F^{\ast}=0$ and $T^{\ast}_{in}=1$. In this case there are four possible steady states depending on whether $T^{\ast}_{nor}$ and $T^{\ast}_{aut}$ are each equal to zero or being positive. If $T^{\ast}_{nor}=T^{\ast}_{aut}=0$, there are two steady states
\[
S^{\ast}_1=\left(0, 0, 1, \dfrac{\lambda_r}{d_r}, 0, 0, 0, 0\right),\quad
S^{\ast}_2=\left(1, 0, 1, \dfrac{\lambda_r}{d_r}, 0, 0, 0, 0\right),
\]
of which $S^{\ast}_1$ is always unstable, and $S^{\ast}_2$ is stable if $cd_F-k\beta>0$, unstable if $cd_F-k\beta<0$, and undergoes a steady-state bifurcation at $cd_F-k\beta=0.$
For $T^{\ast}_{nor}\neq 0$ and $T^{\ast}_{aut}=0$, we again have two steady states
\begin{align*}
&S^{\ast}_3=\left(0, 0, 1, \dfrac{\lambda_r\rho_2}{\rho_2d_r-\rho_1d_n}, \dfrac{d_id_n}{\sigma_1\rho_2}, 0, \dfrac{d_n}{\rho_2}, 0\right),\\
&S^{\ast}_4=\left(1, 0, 1, \dfrac{\lambda_r\rho_2}{\rho_2d_r-\rho_1d_n}, \dfrac{d_id_n}{\sigma_1\rho_2}, 0, \dfrac{d_n}{\rho_2}, 0\right),
\end{align*}
but they are both unstable for any values of parameters.

In the case when $T^{\ast}_{nor}=0$ and $T^{\ast}_{aut}\neq 0$, we have steady states $S^{\ast}_5$ and $S^{\ast}_6$,
\[
\begin{array}{l}
\displaystyle{S^{\ast}_{5}=\left(0, 0, 1,T^{\ast}_{reg},0,\dfrac{d_i\left(d_a+\delta T^{\ast}_{reg}\right)}{\rho_3\sigma_2},\dfrac{d_a+\delta T^{\ast}_{reg}}{\rho_3}, 0\right),}\\\\
\displaystyle{S^{\ast}_{6}=\left(1-\dfrac{\mu_ad_i\left(d_a+\delta T^{\ast}_{reg}\right)}{\rho_3\sigma_2}, 0, 1,T^{\ast}_{reg},0,\dfrac{d_i\left(d_a+\delta T^{\ast}_{reg}\right)}{\rho_3\sigma_2},\dfrac{d_a+\delta T^{\ast}_{reg}}{\rho_3}, 0\right),}
\end{array}
\]
where
\[
\displaystyle{T^{\ast}_{reg}=\frac{d_r\rho_3-\rho_1d_a\pm\sqrt{\left(d_r\rho_3-\rho_1d_a\right)^2-4\rho_1\delta\lambda_r\rho_3}}{2\rho_1\delta}.}
\]
The steady state $S^{\ast}_{5}$ (respectively, $S^{\ast}_{6}$) is stable if the following conditions hold
\begin{equation*}
    \begin{aligned}
	&P<\dfrac{d_a+\delta T^{\ast}_{reg}}{\rho_3}<\dfrac{d_n}{\rho_2},\quad 
	\delta \rho_1\left(T^{\ast}_{reg}\right)^2>\lambda_r\rho_3,\\\\
	&\rho_3\lambda_r^2+\rho_3d_i\lambda_rT^{\ast}_{reg}-\rho_3d_id_a\left(T^{\ast}_{reg}\right)^2-\delta (\rho_1d_a+\rho_3d_i)\left(T^{\ast}_{reg}\right)^3-\rho_1\delta^2\left(T^{\ast}_{reg}\right)^4>0,
	\end{aligned}
\end{equation*}
where
\[
P=\begin{cases}
\dfrac{\sigma_2}{\mu_a d_i},&\mbox{ for }S^{\ast}_{5},\\\\
\dfrac{\sigma_2\left(\beta k-cd_F\right)}{\mu_a d_i\left(c+\beta k\right)},&\mbox{ for }S^{\ast}_{6}.
\end{cases}
\]
This steady state undergoes a steady-state bifurcation if
\begin{align*}
\dfrac{d_a+\delta T^{\ast}_{reg}}{\rho_3}=P,\quad \mbox{or}\quad \dfrac{d_a+\delta T^{\ast}_{reg}}{\rho_3}=\dfrac{d_n}{\rho_2},\quad \mbox{or}\quad \delta \rho_1\left(T^{\ast}_{reg}\right)^2=\lambda_r\rho_3,
\end{align*}
and a Hopf bifurcation if
\begin{equation*}
    \begin{aligned}
	&P<\dfrac{d_a+\delta T^{\ast}_{reg}}{\rho_3}<\dfrac{d_n}{\rho_2},\quad \delta \rho_1\left(T^{\ast}_{reg}\right)^2>\lambda_r\rho_3,\\\\
	&\rho_3\lambda_r^2+\rho_3d_i\lambda_rT^{\ast}_{reg}-\rho_3d_id_a\left(T^{\ast}_{reg}\right)^2-\delta (\rho_1d_a+\rho_3d_i)\left(T^{\ast}_{reg}\right)^3-\rho_1\delta^2\left(T^{\ast}_{reg}\right)^4=0.
	\end{aligned}
\end{equation*}

The steady state with $T^{\ast}_{nor}\neq 0$ and $T^{\ast}_{aut}\neq 0$ only exists for a particular combination of parameters, namely, when
\[
\delta \rho_2^2\lambda_r=(\rho_3d_n-\rho_2d_a)(\rho_2d_r-\rho_1d_n),
\]
and is always unstable.

When $V^{\ast}\neq 0$, all other state variables are also non-zero. In this case, the steady state $S^{\ast}_7$ has $T^{\ast}_{nor}$ and $T^{\ast}_{aut}$ satisfying the following system of equations
\begin{equation*}
\begin{array}{l}
\alpha c\mu_a\rho_2\sigma_2(\beta k+c)(T^{\ast}_{aut})^2T^{\ast}_{nor}+\alpha c\rho_2(\beta k\mu_a\sigma_1+c\mu_F\sigma_2+c\mu_a\sigma_1)T^{\ast}_{aut}(T^{\ast}_{nor})^2\\
-(\alpha\beta ckd_id_n\mu_a+\beta^2k^2d_{in}\rho_2\sigma_2+\alpha\beta ck\rho_2\sigma_2-\alpha c^2d_F\rho_2\sigma_2+\alpha c^2d_id_n\mu_a)T^{\ast}_{aut}\\
T^{\ast}_{nor}+\alpha cd_id_{in}\mu_ap_2(\beta k+c)T^{\ast}_{aut}+\alpha c^2\mu_F\rho_2\sigma_1(T_{nor})^3-(\beta^2k^2d_{in}\rho_2\sigma_1\\
+\alpha\beta ck\rho_2\sigma_1-\alpha c^2d_F\rho_2\sigma_1+\alpha c^2d_id_n\mu_F)(T^{\ast}_{nor})^2+d_i(\alpha c^2d_{in}\mu_Fp_2\\
+\beta^2k^2d_{in}d_n+\alpha\beta ckd_n-\alpha c^2d_Fd_n)T^{\ast}_{nor}-\alpha cd_id_{in}p_2(cd_F-k\beta)=0,\\\\
p_{2}\rho_{1}\rho_{3}{\sigma_{2}}^{2}(T^{\ast}_{aut})^{3}+\sigma_2(-\delta d_{i}p_{1}\rho_{2}+p_{1}\rho_{1}\rho_{2}{\sigma_{2}}+p_{2}\rho_{1}\rho_{2}{\sigma_{2}
}+2p_{2}\rho_{1}\rho_{3}\sigma_{1}\\
-\rho_{1}\rho_{2}{\sigma_{2}}) (T^{\ast}_{aut})^{2}T^{\ast}_{nor} -d_ip_2\sigma_2(d_{a}\rho_{1}+d_{r}\rho_{3}) (T^{\ast}_{aut})^{2}+\sigma_1(-\delta d_{i}p_{1}\rho_{2}\\
+2p_{1}\rho_{1}\rho_{2}\sigma_{2}+2p_{2}\rho_{1}\rho_{2}\sigma_{2}+p_{2}\rho_{1}\rho_{3}{\sigma_{1}}-2\rho_{1}\rho_{2}\sigma_{2})T^{\ast}_{aut}(T^{\ast}_{nor})^{2}+d_i(\delta{d_{i}}d_{n}p_{1}\\
-d_{a}p_{2}\rho_{1}\sigma_{1}-d_{n}p_{1}\rho_{1}\sigma_{2}-d_{n}p_{{2}}\rho_{1}\sigma_{2}-d_{r}p_{1}\rho_{2}\sigma_{2}-d_{r}p_{2}\rho_{2}\sigma_{2}-d_{r}p_{2}\rho_{3}\sigma_{1}\\
+d_{n}\rho_{1}\sigma_{2}+d_{r}\rho_{2}\sigma_{2})T^{\ast}_{aut}T^{\ast}_{nor}+p_2{d_{i}}^{2}(\delta\lambda_{r}+d_{a}d_{r}) T^{\ast}_{aut}+(1-p_{1}-p_{2})T^{\ast}_{nor}\\
\left[-\rho_{1}\rho_{2}{\sigma_{1}}^{2}(T^{\ast}_{nor})^{2}+d_i\sigma_1(d_n\rho_1+d_r\rho_2)T^{\ast}_{nor}-{d_i}^2d_nd_r\right]=0,
\end{array}
\end{equation*}
with the rest of state variables being given by
\begin{align*}
	&I^{\ast}=\dfrac{\sigma_1T^{\ast}_{nor}+\sigma_2T^{\ast}_{aut}}{d_i}, \quad A^{\ast}=\dfrac{c\left(d_F+\mu_FT^{\ast}_{nor}+\mu_aT^{\ast}_{aut}\right)}{k\beta}, \quad V^{\ast}=\dfrac{1-A^{\ast}-\mu_aT^{\ast}_{aut}}{\beta},\\
	&F^{\ast}=\dfrac{cV^{\ast}}{k}, \quad T^{\ast}_{in}=\dfrac{d_{in}}{d_{in}+\alpha F^{\ast}}, \quad T^{\ast}_{reg}=\dfrac{\lambda_r+p_1\alpha T^{\ast}_{in}F^{\ast}}{d_r-\rho_1I^{\ast}}.
\end{align*}
It does not prove possible to analyse stability of this steady state analytically, hence, one has to resort to numerical calculations.\\

\noindent {\bf Remark.} Inclusion of a term corresponding to production of autoreactive T cells directly from normal activated T cells in a manner similar to Blyuss and Nicholson \cite{blyu12,blyu15} would make the steady states $S_3^*$ and $S_4^*$ infeasible, while having no major effect on stability of other steady states. Hence, it suffices to consider the above model without explicitly modelling the transition from $T_{nor}$ to $T_{aut}$.

\begin{table}[h]
	\label{parameter table}
	{\begin{tabular}{l l l}
			\toprule
			Parameter & Value & Definition \\
			\hline
			$\beta$ & 3 & Infection rate \\
			\hline
			$\mu_a$ & 20 & The rate of killing of uninfected cells by autoreactive T cells\\
			\hline
			$d_F$ & 1.1 & Natural death rate of infected cells\\
			\hline
			$\mu_F$ & 6 & The rate of killing of infected cells by the normal T cells\\
			\hline
			$d_{in}$ & 1 & Growth rate of na\"ive T cells\\
			\hline
			$\alpha$ & 0.4 &  Rate of activation of na\"ive T cells by infected cells\\
			\hline
			$\lambda_r$ & 3 & Growth rate of regulatory T cells\\
			\hline
			$d_r$ & 0.4 & Natural death rate of regulatory T cells\\
			\hline
			$p_1$ & 0.4 & Rate of conversion of na\"ive T cells into regulatory T cells\\
			\hline
			$p_2$ & 0.4 & Rate of conversion of na\"ive T cells into normal T cells\\
			\hline
			$\rho_1$ & 10 & Proliferation rate of regulatory T cells by interleukin 2 (IL-2) \\
			\hline
			$\rho_2$ & 0.8 & Proliferation rate of normal T cells by interleukin-2 (IL-2)\\
			\hline
			$\rho_3$ & 2 & Proliferation rate of autoreactive T cells by interleukin 2 (IL-2)\\
			\hline
			$d_n$ & 1 & Natural death rate of normal T cells\\
			\hline
			$d_a$ & 0.001 & Natural death rate of autoreactive T cells\\
			\hline
			$\delta$ & 0.002 & Rate of clearance of autoreactive T cells by regulatory T cells\\
			\hline
			$\sigma_1$ & 0.15 & Rate of production of interleukin-2 (IL-2) by normal T cells\\
			\hline
			$\sigma_2$ & 0.2 & Rate of production of interleukin-2 (IL-2) by autoreactive T cells\\
			\hline
			$d_i$ & 0.6 & Natural clearance rate of IL-2\\
			\hline
			$k$ & 2 & Rate of production of free virus\\
			\hline
			$c$ & 6 & Natural clearance rate of free virus\\
			\hline
	\end{tabular}}
	\caption{Table of parameters}
\end{table}

\section{Numerical stability analysis and simulations}
\label{sec:4}

\begin{figure}[h]
	\includegraphics[width=1\linewidth]{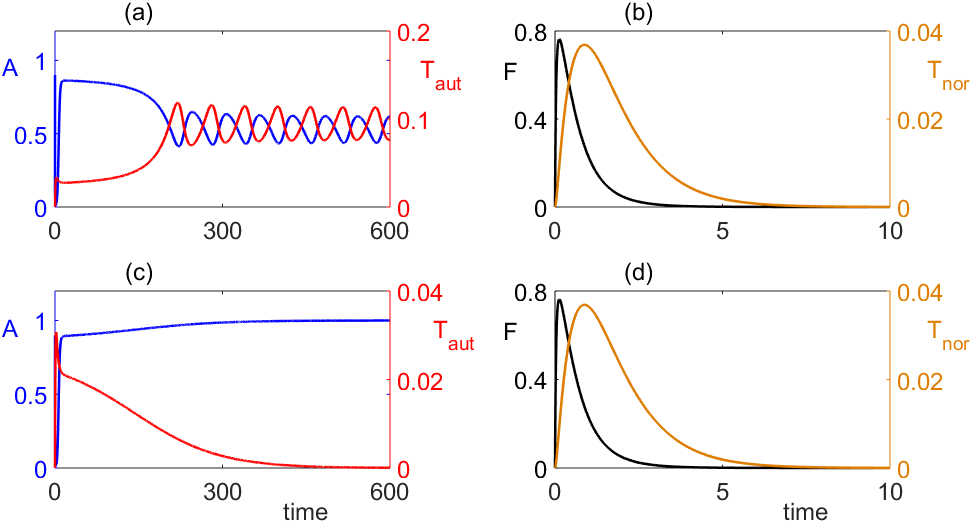}
	\caption{Simulation of the model (\ref{resc_model}) with parameter values from Table 1, except for $\lambda_r=2.5$, $\mu_a=5$, $\mu_F=10$, and the initial condition (\ref{IC}). (a), (b) $V(0)=10$, $T_{reg}(0)=100$, the system exhibits periodic behaviour around $S^{\ast}_6$, i.e. clearance of infection followed by the onset of autoimmune response. (c), (d) $V(0)=10$, $T_{reg}(0)=150$, the model converges to a stable disease-free steady state $S^{\ast}_2$.}
	\label{simulation1}
	\vspace{-0.3cm}
\end{figure}

To investigate various dynamical scenarios that can be exhibited by the model, we now perform a comprehensive numerical analysis of stability of different steady states and identify their possible bifurcations. Analytical results from the previous section suggest that the disease-free steady state $S^{\ast}_2$ is stable when $k\beta<cd_F$, and unstable when $k\beta>cd_F$. As will be shown below, there are some major differences in dynamics between these two parameter combinations, hence, we will analyse them separately. Since there are significant differences in the reported values of many of the model parameters, and some of them have not yet been properly measures, we fix the baseline values as given in Table 1, and perform a sweep of parameter space to identify the effects of varying these parameters. Since prior to the start of infection, the numbers of infected cells, normal activated T cells, and the amount of IL-2 are all equal to zero, the initial condition for the model is taken to be
\begin{equation}\label{IC}
	(A(0),F(0),T_{in}(0),T_{reg}(0),T_{nor}(0),T_{aut}(0),I(0),V(0))=(0.9,0,0.8,0.7,0,0,0,0.4),
\end{equation}
which indicates the presence of some number of free virus particles. Here the initial values of $A(0)$ and $T_{in}(0)$ are chosen randomly, with the only requirement that they do not exceed unity, in light of the fact that we are considering a non-dimensionalised model. For analysis of basins of attraction, the values of $T_{reg}(0)$ and $V(0)$ will be varied.
\begin{figure}[h]
	\centering
	\includegraphics[width=\linewidth]{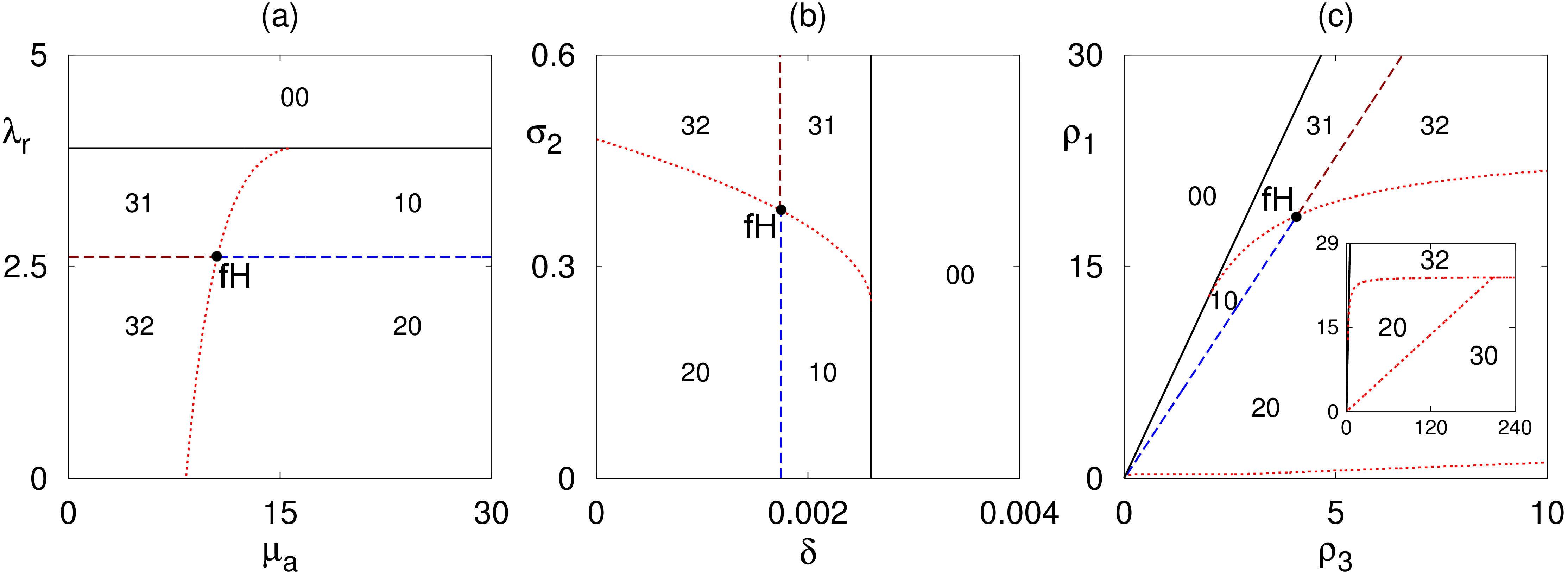}
	\caption{Regions of feasibility and stability of the steady states $S^{\ast}_5$ and $S^{\ast}_6$ with parameter values from Table 1. Black and red curves indicate the boundaries of feasibility and the steady-state bifurcation, whereas dashed lines (blue/brown) show the boundaries of Hopf bifurcation of the steady states $S^{\ast}_5$ and $S^{\ast}_6$, respectively, with `fH' indicating the fold-Hopf bifurcation. The first digit of the index refers to $S^{\ast}_5$, while the second corresponds to $S^{\ast}_6$, and they indicate that in that parameter region the respective steady state is unfeasible (index is `0'), stable (index is `1'), unstable via Hopf bifurcation with a periodic orbit around this steady state (index is `2'), or unstable via a steady-state bifurcation (index is `3'). In all plots, the condition $k\beta<cd_F$ holds, so the disease-free steady state $S^{\ast}_2$ is also stable.}
	\vspace{-0.5cm}
	\label{S5S6}
\end{figure}
Figure~\ref{S5S6} illustrates how system dynamics is affected by the parameters. Since the condition $k\beta<cd_F$ holds, the disease-free steady state $S^{\ast}_2$ is always stable. However, the system can also have two other biologically feasible steady states $S^{\ast}_5$ and $S^{\ast}_6$, which only exist, provided regulatory T cells do not grow too rapidly and do not clear autoreactive T cells too quickly. In the case where autoreactive T cells are very effective in killing infected cells (i.e. for higher $\mu_a$), or when they are producing IL-2 at a slow rate (smaller $\sigma_2$), only the steady state $S^{\ast}_5$ is feasible, which has the zero population of host cells $A$, while the steady state $S^{\ast}_6$ with $A>0$ can only exist when $\mu_a$ is relatively low (or $\sigma_2$ is high), and $S^{\ast}_5$ is unstable. Provided the steady states $S^{\ast}_5$ and $S^{\ast}_6$ are feasible, decreasing the growth rate $\lambda_r$ of regulatory T cells results in a supercritical Hopf bifurcation, which gives rise to stable periodic solutions around these steady states. Since the steady state $S^{\ast}_5$ is characterised by $A=0$, both regimes where this steady state is stable, or unstable with oscillations around it, biologically correspond to a situation where the host cells are dead. On the other hand, a periodic solution around $S^{\ast}_6$ corresponds to a proper autoimmune response, whereby the infection is cleared, but the immune system still exhibits endogenous oscillations, as illustrated in Fig.~\ref{simulation1}(a)-(b). At the intersection of the lines of Hopf bifurcation and the steady-state bifurcation, one has the fold-Hopf (also known as zero-Hopf or saddle-node Hopf) bifurcation \cite{kuz98}. Importantly, the steady states $S^{\ast}_5$ and $S^{\ast}_6$ can only exist if the rate $\rho_3$ at which IL-2 promotes proliferation of autoreactive T cells is sufficiently high, and this minimum value of the rate $\rho_3$ increases linearly with the rate $\rho_1$ at which IL-2 promotes proliferation of regulatory T cells. Once feasible, the steady states $S^{\ast}_5$ and $S^{\ast}_6$ are stable for smaller values of $\rho_3$ and then undergo Hopf bifurcation, when $\rho_3$ is sufficiently increased.

\begin{figure}[h]
\begin{center}
	\includegraphics[width=0.85\textwidth]{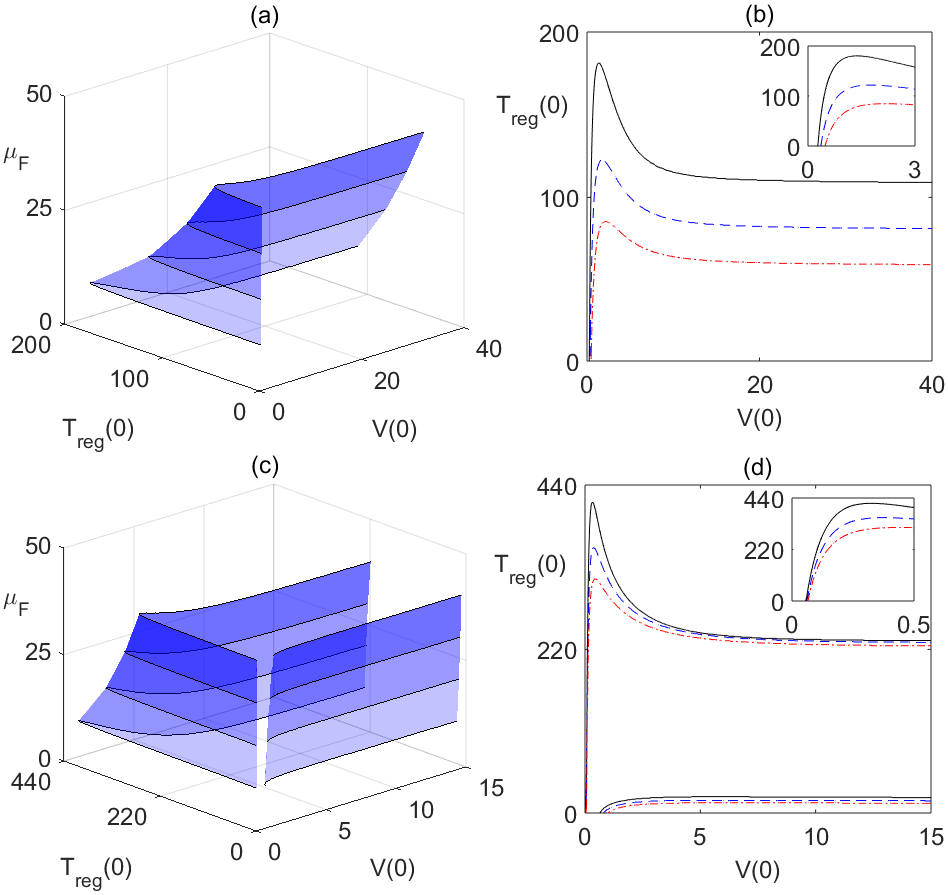}
	\caption{Regions of bi-stability with parameter values from Table 1 and initial condition (\ref{IC}). (a), (b) $\lambda_r=2.5$, $\mu_a=5$, with $\mu_F=10$ (black), $\mu_F=20$ (blue), $\mu_F=30$ (red). The system exhibits autoimmune response to the right of the surface in (a) and below the curves in (b), while to the left of the surface in (a) and above the curves in (b) it tends to a stable disease-free steady state $S^{\ast}_2$. (c), (d) $\rho_1=30$, $\rho_3=8$. The system exhibits autoimmune response inside the region bounded by the surfaces in (c), or by the curves in (d), and outside it tends to a stable disease-free steady state $S^{\ast}_2$.}
	\end{center}
	\vspace{-0.3cm}
	\label{bistability1}
\end{figure}

Since for all parameter combinations in Fig.~\ref{S5S6} the steady state $S^{\ast}_2$ is stable, this means that the system can exhibit bi-stability between steady states and/or periodic solutions. To investigate this in more detail, we choose parameter values in the `32' region in Fig.~\ref{S5S6}, where periodic oscillations around the steady state $S^{\ast}_6$ are possible. While previous work on multi-stability in models of autoimmunity focussed mainly on identifying parameter regions associated with bi-stability \cite{And15,Bak13,Li17}, the structure of basins of attraction associated with different dynamical states has remained largely unexplored. To analyse basins of attraction in our model, due to high dimensionality of the phase space, we fix initial conditions for all state variables, and consider different initial amounts of free virus $V(0)$ and regulatory T cells $T_{reg}(0)$, as illustrated in Fig.~\ref{bistability1}. Biologically, this corresponds to varying the initial level of infection, as well as the initial state of the immune system, which can be primed by previous exposures to other pathogens. This Figure shows that if the initial number of regulatory T cells is sufficiently high, the system is able to successfully eliminate infection without any lasting consequences, settling on a stable disease-free steady state. Interestingly, for very small initial amounts of free virus, a higher amount of regulatory T cells is required to clear the infection. For lower values of $T_{reg}(0)$, the system exhibits stable periodic oscillations around the steady state $S^{\ast}_6$, which biologically represents the regime of autoimmune response. One can also observe that the minimum value of $T_{reg}(0)$ needed to achieve a diease-free steady state reduces with increasing the rate $\mu_F$ at which normal T cells are able to kill infected cells. Figure~\ref{simulation1} illustrates temporary evolution of the system in the regime of bi-stability between a stable disease-free steady state and a periodic solution, corresponding to autoimmunity. The dynamics of regulatory T cells (not shown in this figure) mimics that of autoreactive T cells.

\begin{figure}[h]
	\includegraphics[width=1\linewidth]{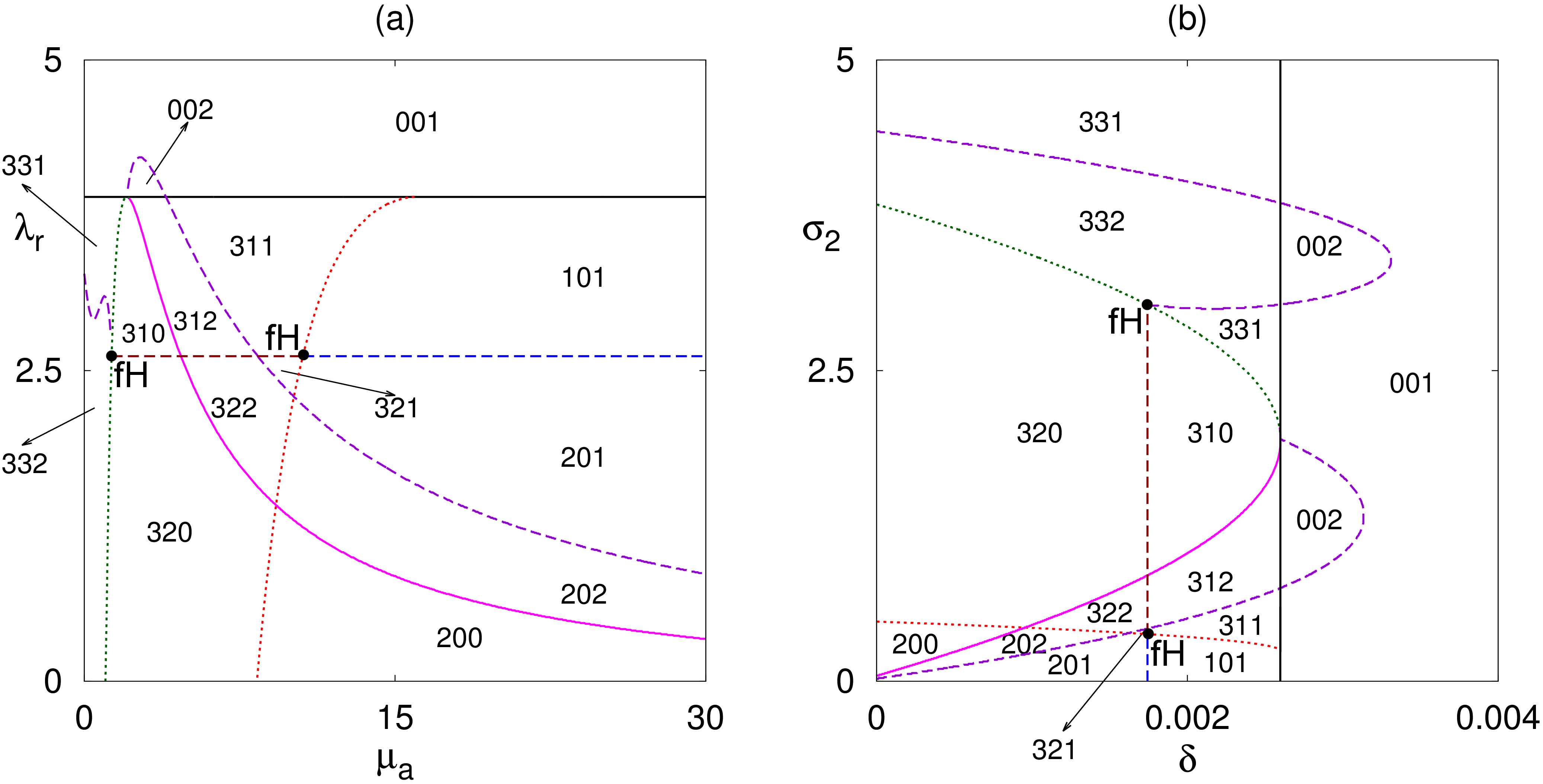}
	\caption{Regions of feasibility and stability of the steady states $S^{\ast}_5$, $S^{\ast}_6$, and $S^{\ast}_7$ with parameter values from Table 1. Black and magenta curves indicates the boundaries of feasibility for $S^{\ast}_5/S^{\ast}_6$ and $S^{\ast}_7$, dashed curves are the boundaries of Hopf bifurcation for $S^{\ast}_5/S^{\ast}_6$ (blue/brown) or $S^{\ast}_7$ (purple), and dotted lines are the boundaries of the steady-state bifurcation of $S^{\ast}_5$ (red) and $S^{\ast}_6$ (green), with `fH' indicating the location of the fold-Hopf bifurcation. The first digit of the index refers to $S^{\ast}_5$, the second corresponds to $S^{\ast}_6$, and the third corresponds to $S^{\ast}_7$. These indices indicate that in that parameter region the respective steady state is unfeasible (index is `0'), stable (index is `1'), unstable via Hopf bifurcation with a periodic orbit around this steady state (index is `2'), or unstable via a steady-state bifurcation (index is `3'). In all plots, the condition $k\beta>cd_F$ holds, so the disease-free steady state $S^{\ast}_2$ is unstable.}
	\label{SSS}
\end{figure}

Next, we consider a situation described by the combination of parameters satisfying $k\beta>cd_F$, so the disease-free steady state $S^{\ast}_2$ is unstable, and the system can only have steady states $S^{\ast}_5$, $S^{\ast}_6$, and $S^{\ast}_7$. Figure~\ref{SSS} shows how feasibility and stability of these steady states depend on parameters. Naturally, this figure is identical to Fig.~\ref{S5S6} in terms of indicating stability and bifurcations of the steady states $S^{\ast}_5$ and $S^{\ast}_6$. One should note that unlike the case considered earlier, now for sufficiently high rate $\sigma_2$ of production of IL-2, or sufficiently small rate $\mu_a$ at which autoreactive T cells are killing infected
\newpage
\noindent cells, the steady state $S^{\ast}_6$ can also undergo a steady-state bifurcation due to the fact that the condition $k\beta>cd_F$ holds. Beyond this stability boundary, i.e. for very high values of $\sigma_2$ or very small values of $\mu_a$, both steady states $S^{\ast}_5$ and $S^{\ast}_6$ are unstable, and the system settles either on the steady state $S^{\ast}_7$, or an a periodic solution around this steady state. In the parameter region, where only the steady state $S^{\ast}_7$ is feasible, this steady state can only be unstable, giving rise to stable periodic oscillations, for sufficiently small values of $\delta$ or $\lambda_r$, whereas for higher values of those parameters this steady state is stable.

\begin{figure}
	\includegraphics[width=1\linewidth]{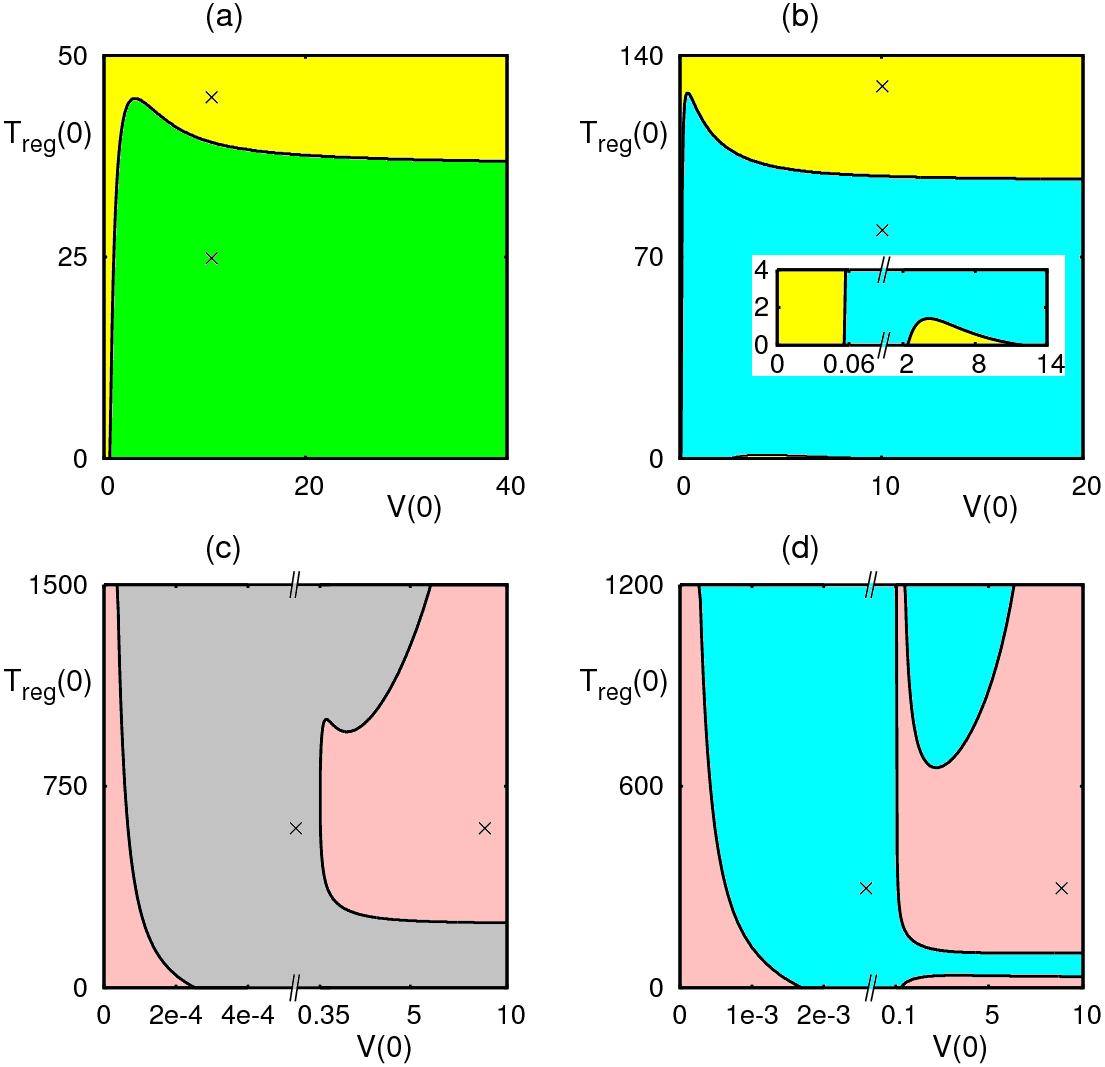}
	\caption{Regions of bi-stability in the system (\ref{resc_model}) with the initial condition (\ref{IC}), parameter values from Table 1, and $\beta=4$, $k=2.1$. (a) $\delta=0.002$, $\sigma_2=0.2$, (b) $\delta=0.0024$, $\sigma_2=0.5$, (c) $\delta=0.0017$, $\sigma_2=0.42$, (d) $\delta=0.0024$, $\sigma_2=0.7$. Green, blue and yellow are the basins of attraction of the steady states $S^{\ast}_5$, $S^{\ast}_6$, and $S^{\ast}_7$, respectively. Grey and pink are the basins of attraction of periodic solutions around $S^{\ast}_6$ and $S^{\ast}_7$, respectively.}
	\label{bistability2}
	\vspace{-0.5cm}
\end{figure}

Figure \ref{bistability2} demonstrates the basins of attraction of the steady states $S^{\ast}_5$, $S^{\ast}_6$ and $S^{\ast}_7$, as well as periodic solutions around them. As it has already been mentioned, this is the first time when basins of attraction for different steady states and periodic solutions are identified in a model of cytokine-mediated immune response and autoimmunity. Figure (a) shows that if the initial amount of free
\newpage
\noindent virus is sufficiently small, the system will converge to $S^{\ast}_7$ for any value of $T_{reg}(0)$. For higher values of $V(0)$, the system exhibits bi-stability, where for smaller initial numbers of regulatory T cells $T_{reg}(0)$ it converges to the stable steady state $S^{\ast}_5$ corresponding to the death of susceptible organ cells, while for higher values of $T_{reg}(0)$, the system settles on the stable steady state $S^{\ast}_7$. While the critical value of $T_{reg}(0)$ at which the transition between the two steady state takes place initially increases with $V(0)$, eventually it settles on some steady level, so that for higher initial amounts of free virus, this critical value no longer depends on $V(0)$. Figure (b) illustrates a qualitatively similar behaviour for higher rates of production of IL-2 and clearance of autoreactive T cells, in which case there is a bi-stability between $S^{\ast}_6$ and $S^{\ast}_7$, but with the difference that there is also a small region for small values of $T_{reg}(0)$ and intermediate values of $V(0)$, where the system also converges to $S^{\ast}_7$. Figures (c) and (d) illustrate bi-stability between a periodic solution around $S^{\ast}_7$ and either the stable steady state $S^{\ast}_6$, or a periodic solution around this steady state.

\begin{figure}[ht]
	\vspace{-0.3cm}
	\includegraphics[width=1\linewidth]{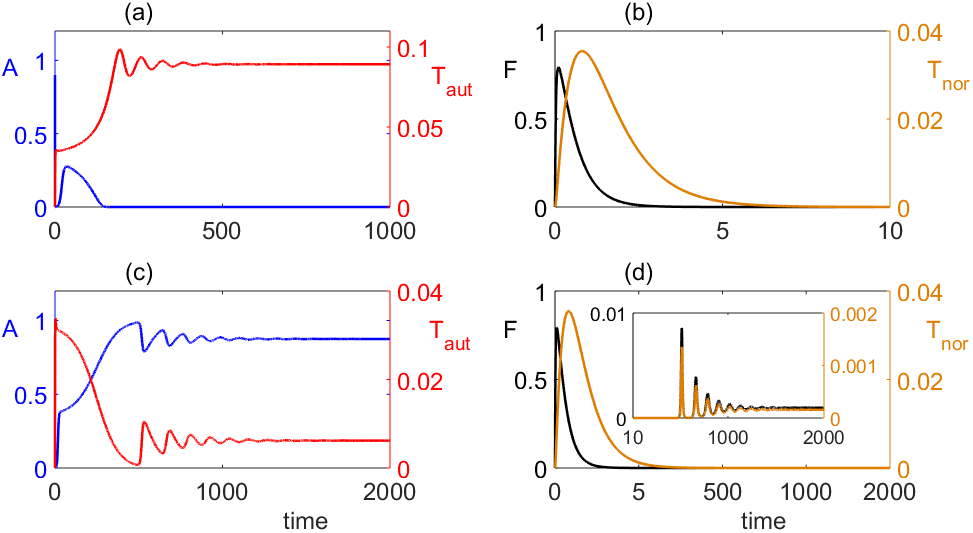}
	\caption{Numerical solution of the model (\ref{resc_model}) with the initial condition (\ref{IC}), parameter values from Table 1, and $\beta=4$, $k=2.1$, $\delta=0.002$, $\sigma_2=0.2$. In (a) and (b) $V(0)=10$ and $T_{reg}(0)=25$. The model converges to the $S^{\ast}_5$. In (c) and (d) $V(0)=10$ and $T_{reg}(0)=45$. The model converges to the $S^{\ast}_7$. The dynamic of $T_{reg}$ has a same behaviour as $T_{aut}$.}
	\label{simulation2}	
\end{figure}

\begin{figure}
	\includegraphics[width=1\linewidth]{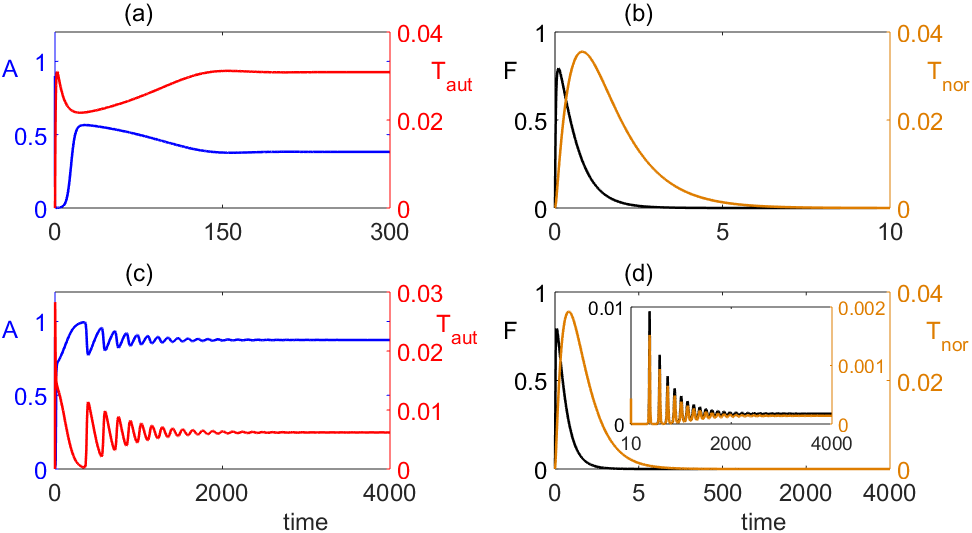}
	\caption{Numerical solution of the model (\ref{resc_model}) with the initial condition (\ref{IC}), parameter values from Table 1, and $\beta=4$, $k=2.1$, $\delta=0.0024$, $\sigma_2=0.5$. (a), (b) $V(0)=10$, $T_{reg}(0)=80$. The system converges to $S^{\ast}_6$. (c), (d) $V(0)=10$, $T_{reg}(0)=130$. The system converges to $S^{\ast}_7$. The dynamics of $T_{reg}$ is the same as $T_{aut}$.}
	\label{simulation5}
	\vspace{-0.3cm}
\end{figure}

\begin{figure}
	\centering
	\includegraphics[width=\linewidth]{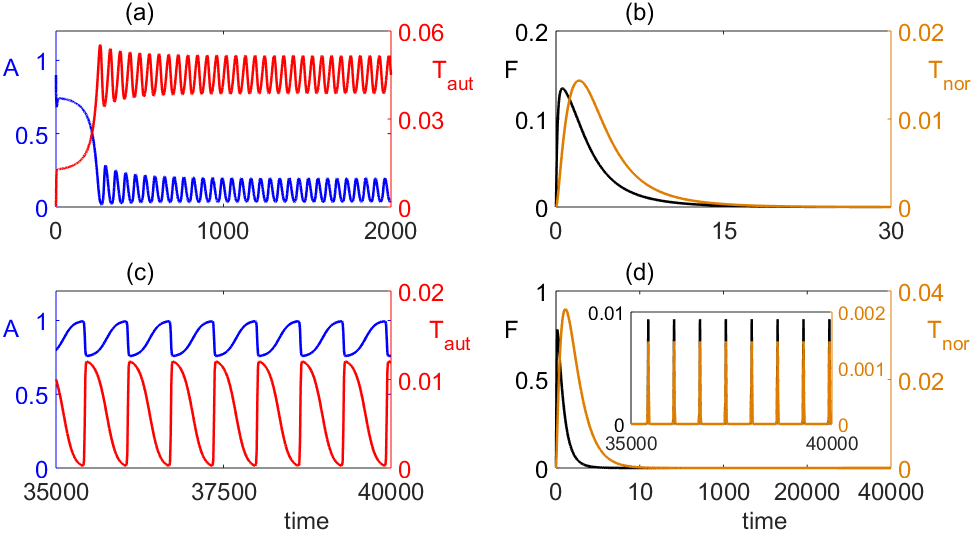}
	\caption{Numerical solution of the model (\ref{resc_model}) with the initial condition (\ref{IC}), parameter values from Table 1, and $\beta=4$, $k=2.1$, $\delta=0.0017$, $\sigma_2=0.42$. (a), (b) $V(0)=0.3$, $T_{reg}(0)=600$. This system exhibits periodic oscillations around $S^{\ast}_6$, i.e. an autoimmune response. (c), (d) $V(0)=9$, $T_{reg}(0)=600$. The system exhibits periodic oscillations around $S^{\ast}_7$. The dynamics of $T_{reg}$ is the same as $T_{aut}$.}
	\label{simulation3}
	\vspace{-0.3cm}
\end{figure}

\begin{figure}
	\centering
	\includegraphics[width=\linewidth]{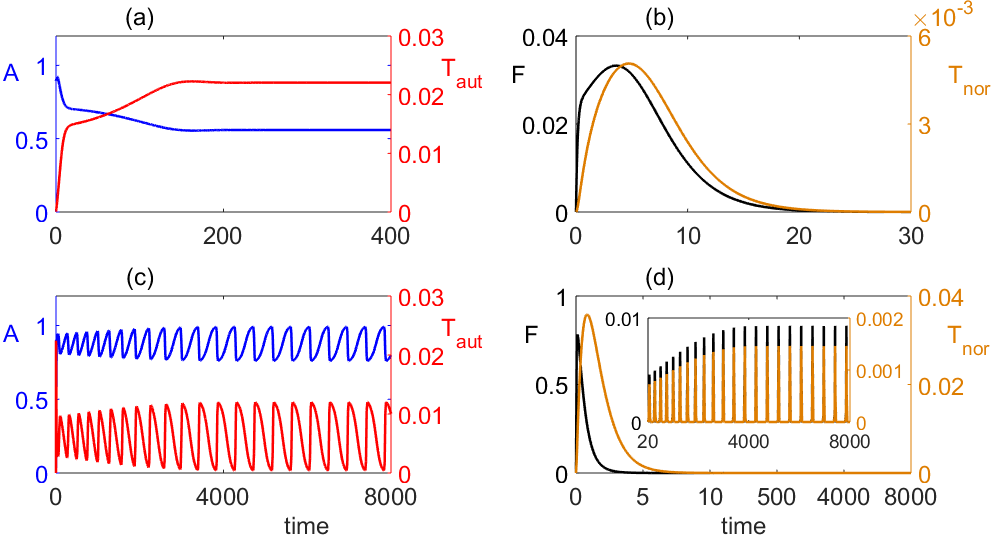}
	\caption{Numerical simulation of the model (\ref{resc_model}) with the initial condition (\ref{IC}), parameter values from Table 1, and $\beta=4$, $k=2.1$, $\delta=0.0024$, $\sigma_2=0.7$. (a), (b) $V(0)=0.05$, $T_{reg}(0)=300$. The system converges to $S^{\ast}_6$. (c), (d) $V(0)=9$, $T_{reg}(0)=300$. The system exhibits periodic oscillations around $S^{\ast}_7$. The dynamics of $T_{reg}$ is the same as $T_{aut}$.}
	\label{simulation4}
	\vspace{-0.3cm}
\end{figure}

Numerical simulations in Figs.~\ref{simulation2}, \ref{simulation5}, \ref{simulation3}, and \ref{simulation4} show the dynamics of the model in the case when $k\beta>cd_F$ for the same parameter values but different initial conditions, thus illustrating various bi-stability scenarios shown in Fig.~\ref{bistability2}, in which crosses indicate the values of specific initial conditions used for simulations. Figure~\ref{simulation2} demonstrates how for sufficiently small initial number of regulatory T cells the infection can result in the death of organ cells, in which case the system approaches a stable steady state $S^{\ast}_5$. On the other hand, for a higher number of Tregs, the system goes to a stable steady state $S^{\ast}_7$ which represents a persistent (chronic) infection. In this case, one observes some kind of balance maintained with the help of regulatory T cells: while the immune system is not able to clear the infection, at the same time it prevents infection from destroying the organ cells.

Figure~\ref{simulation5} illustrates a similar behaviour, where bi-stability takes place between the steady states $S^{\ast}_6$ and $S^{\ast}_7$. In this case, for a smaller number of regulatory T cells, the system favours the regime
\newpage
\noindent of normal clearance of infection, where after the initial growth, the numbers of infected cells and activated T cells responding to foreign antigen go to zero. For higher numbers of regulatory T cells, the system again approaches a stable steady state $S^{\ast}_7$ describing a persistent infection. This is a really interesting and counter-intuitive result, which suggests that whilst regulatory T cells play a major role in reducing autoimmune response during normal disease clearance, when they are present in high numbers, the are actually promoting the persistence of infection.

Figure~\ref{simulation3} illustrates a regime of bi-stability between periodic solutions around the steady states $S^{\ast}_6$ and $S^{\ast}_7$. Similarly to the case of stable disease-free steady state considered earlier, the periodic solution around $S^{\ast}_6$ biologically corresponds to the regime of autoimmune response, where upon clearance of the initial infection, the immune systems maintains endogenous oscillations, in which the growth of autoreactive T cells results in the destruction of some healthy organ cells, after which the number of autoreactive T cells

\noindent decreases, and the organ cells recover.

One should note that since these oscillations take place around the steady state $S^{\ast}_6$, the mean concentration of organ cells is much lower than what it was before the infection. In the case of periodic oscillations around the steady state $S^{\ast}_7$, initially one observes a similar behaviour in terms of rapid growth of infected cells, followed by an expansion in the population of activated T cells recognising foreign antigen, but after the number of infected cells decreases, rather than go to zero, it
\newpage
\noindent settles on periodic oscillations around some small positive level. This suggests that the infection is not cleared, but rather than being chronic, there are intervals of quiescence where the level of infection is very small, followed by regular intervals of rapid growth of infection and autoreactive T cells, which causes significant reduction in the number of uninfected organ cells. After this infection is significantly reduced by the activated T cells, the cycle repeats.

Finally, Figure~\ref{simulation4} demonstrates a situation where the system has a bi-stability between a stable steady state $S^{\ast}_6$ and a periodic solution around $S^{\ast}_7$. The difference from the previous case is that instead of autoimmune regime, the system can now successfully clear the infection, without having any subsequent oscillations. Although the infection itself is cleared, it leaves an imprint on the dynamics in the form of a reduced number of organ cells and a non-zero number of autoreactive T cells.

\section{Discussion}
\label{sec:5}

In this paper we have developed and studied a model of immune response to a viral infection, with an emphasis on the role of cytokine mediating T cell activity, and T cells having different activation thresholds. Stability analysis of the model's steady states has allowed us to identify regimes with different dynamical behaviour depending on system parameters. When the product of infection rate and the rate of production of new virus particles is smaller than the product of the rates of viral clearance and death of infected cells, the immune system is able to successfully clear the infection without further consequences for either the host organ cells, or the immune system. In this case, the system settles on a stable disease-free steady state, characterised by the absence of infected cells and free virus, as well as zero amount of normal or autoreactive T cells. Another biologically feasible steady state that can exist in some parameter regimes is the state that also has no infected cells or free virus, but maintains non-zero levels of activated T cells. We have derived analytical conditions for steady-state and Hopf bifurcations of this state. When the disease-free steady state is unstable, the model also possesses a steady state with all cell populations being positive, which biologically corresponds to a state of chronic infection.

To investigate how the system behaves in different parameter regimes, we have solved it numerically, paying particular attention to cases where more than one steady state can be feasible. This has allowed us to identify regions of multi-stability, where for the same parameter values, depending on the initial conditions the system can approach either two distinct steady states, or a steady state and a periodic solution. In the case where the disease-free steady state is stable, such a regime has a very important potential clinical significance, as effectively it suggests that whether or not a given patient is able to clear the infection or will go on to develop autoimmunity depends not only on the rate of performance of their immune system, but also on the magnitude of viral challenge they experience and the amount of regulatory T cells they have before the infection. Numerical simulations for the autoimmune regime illustrate how initial infection leads to a rapid growth in the number of infected cells, resulting in the growth of populations of normal and autoreactive T cells, which clear the infection, but on a longer time-scale the system exhibits sustained periodic oscillations that can be associated with periods of relapses and remission, characteristic for many autoimmune diseases. For the case where the disease-free steady state is unstable, the bi-stability can occur between an autoimmune steady state and a chronic state, or a period orbit around the latter. A number of earlier models have looked into bi-stability in the immune dynamics, and the model analysed in this paper provides further clues regarding the important role played by cytokines in controlling the dynamics of immune response. In the regime of bi-stability, we have discovered that the initial state of the immune system, and the initial viral load determine the course and outcome of the immune response, a result that would be interesting to test in an experimental setting.

There are several directions in which the model presented in this paper could be extended to make it more realistic. One possibility is to include in the model other potentially relevant aspects of immune system dynamics, such as memory T cells \cite{Skap05,Antia05}, or the effects of T cells on secretion of IL-2 \cite{burr2,Fat18b}. Another aspect that is particularly relevant for our model is the fact that activation thresholds can themselves change during the process of immune response, hence, one could explicitly include the dynamics of activation thresholds as an extra component of the model \cite{Gsinger96,Bonn05,berg,Scher04}. Many viruses are known to have a non-negligible lag phase in their virus cycle, which includes such processes as virus attachment, cell penetration and uncoating, virus assembly, maturation, and release of new virions. All these processes result in the delay in production and release of new virus particles, thus having an impact on the dynamics of immune response and potential onset and development of autoimmune disease. Mathematically, the lag phase can be represented using time delays in the relevant terms of the model, and the available data on lag phase for viruses associated with triggering or exacerbating autoimmune diseases can be used to validate the model. 

Due to the large dimension of the phase space of model (\ref{resc_model}), and the complex structure of basins of attraction for different steady states and periodic solutions illustrated above, it would be important to more systematically investigate the issue of multi-stability in the system. One very promising approach to address this challenging problem is that of {\it offset boosting} \cite{Li16a,Li16b,Li17a,Li17b}. The underlying idea is that provided the model has the property that ensures that the functional form of equations remains the same when one of the state variables is shifted by a constant, this would imply that the system is {\it variable-boostable}, and it can then be transformed into a system with {\it conditional symmetry} \cite{Li17b}, implying that the new system is symmetric subject to a certain condition on the boosting.
The importance of this conditional symmetry is two-fold. On the one hand, it provides a robust handle on identification of symmetric attractors in the system \cite{Li17b}, which otherwise may not be apparent, especially due to a large dimensionality of the phase space. On the other hand, this methodology yields an effective tool for identifying multi-stability and exploring basins of attraction for different dynamical states, by allowing the system to visit different basins of attraction with the same initial condition by means of offset boosting \cite{Li17a}. Future work will explore model (\ref{resc_model}) from the perspective of conditional symmetry to identify all parameter regions associated with multi-stability and to delineate associated basins of attraction. This will provide significant practical insights for monitoring of onset and progress of autoimmune disease.

\section*{Acknowledgments} \noindent FF acknowledges the support from Chancellor's Studentship from the University of Sussex.

\bibliographystyle{ieeetr}
\bibliography{autoimmunity} 

\end{document}